\documentclass[%
 preprint,
 superscriptaddress,
  aps,
 prl,
]{revtex4-2}
\usepackage{amsmath}
\usepackage{textcomp}
\usepackage{graphicx}
\usepackage{xcolor}%
\usepackage{dcolumn}
\usepackage{bm}
\usepackage{caption}

\newcommand\zp[1]{{\color{blue}#1}}

\newcommand\zfig[1]{{\color{gray}#1}}

\begin{document}


\title{Electrified Fracture of Nanotube Films}

\author{Jinbo Bian}
\affiliation{%
Applied Mechanics Laboratory and Department of Engineering Mechanics, Tsinghua University, Beijing, 100084, China.
}%
\affiliation{%
Center for Nano and Micro Mechanics, Tsinghua University, Beijing 100084, China.
}%

\author{Shijun Wang}
\affiliation{%
CAS Key Laboratory of Nanosystem and Hierarchical Fabrication, CAS Center for Excellence in Nanoscience, National Center for Nanoscience and Technology, Chinese Academy of Sciences, Beijing 100190, China.
}%

\author{Zhaokuan Yu}
\affiliation{%
Department of Physics, Zhejiang University, Hangzhou 310027, China.
}
\affiliation{%
Center for Nano and Micro Mechanics, Tsinghua University, Beijing 100084, China.
}

\author{Zhong Zhang}
\email{zhongzhang@ustc.edu.cn}
\affiliation{%
CAS Key Laboratory of Mechanical Behavior and Design of Materials, Department of Modern Mechanics, School of Engineering Science, University of Science and Technology of China, Hefei 230027, China
}%
\affiliation{%
CAS Key Laboratory of Nanosystem and Hierarchical Fabrication, CAS Center for Excellence in Nanoscience, National Center for Nanoscience and Technology, Chinese Academy of Sciences, Beijing 100190, China.
}

\author{Zhiping Xu}
\email{xuzp@tsinghua.edu.cn}
\affiliation{%
Applied Mechanics Laboratory and Department of Engineering Mechanics, Tsinghua University, Beijing, 100084, China.
}%
\affiliation{%
Center for Nano and Micro Mechanics, Tsinghua University, Beijing 100084, China.
}%

\begin{abstract}
Strong and conductive carbon nanotube films are ideal candidates for lightning-strike protection.
Understanding their failure mechanisms by considering the anisotropic and single-fiber nature is essential to improve performance.
Our experimental studies show that the single-layer, nanometer-thick films fail under electrification by crack nucleation and propagation, reminiscent of brittle and ductile fracture of materials under mechanical loads.
Sharp and diffuse patterns of fracture are identified in aligned and non-woven films, respectively, signaling the strong effect of material anisotropy that is absent in common engineering materials.
The fracture is driven by local Joule heating concentrated at the crack fronts instead of force-induced breakage, which is validated by experimental characterization and simulation results at both continuum and atomistic levels.
\end{abstract}

\maketitle


\clearpage
\newpage

Lightning striking is an instantaneous but energy-intensive process.
The current rises to 200 kA within tens of \textmu s \zp{~\cite{larsson2002interaction,kumar2020factors}}, resulting in concentrated damage to materials\zp{~\cite{uman2003interaction,larsson2002interaction,gagne2014lightning}}.
Carbon nanotubes (CNTs) are one-dimensional (1D) nanostructures that are mechanically strong and conductive for both heat and charge carriers\zp{~\cite{dresselhaus1998physical}}.
CNT fibers, films, and composites have thus been explored for their potential in high-end material applications for lightning-strike protection\zp{~\cite{de2013carbon,kinloch2018composites,hirano2010artificial}}.
However, unlike conventional protection materials in use (e.g., copper, aluminum), the performance of CNT-derived materials highly relies on their microstructures, such as the orientation and packing density of CNTs, since the van der Waals contact between CNTs is much weaker than the sp$^{2}$ bonding network within the graphitic walls.
Recent studies showed improved mechanical and conductive properties of aligned CNT films manufactured from vertically-grown arrays compared to their randomly aligned counterparts, as well as promising performance against lightning strikes\zp{~\cite{bai2023mechanism,bai2023superaligned}}.
However, the failure mechanism remains unclear due to both the multiphysical nature of the underlying process and the multiscale structures of the materials.

Among the various types of material damage, fracture behaves as a detrimental process for structural health since the cracks can be nucleated from local energy concentration and quickly advance across the whole structure in a catastrophic way.
Fracture can be caused by different driving forces under lightning strikes, including mechanical loads, thermal stress or decomposition, chemical attack, and electrical breakdown\zp{~\cite{huntington1961current,subramaniam2013one}}.
The characteristics and dynamics of fracture can identify the sources of cracks in damaged materials.
For example, the crack morphologies are often associated with the underlying mechanisms, such as the sharp and blunt ones in overloaded brittle and ductile solids, respectively\zp{~\cite{lawn_2004}}.
In comparison, fractures under electrification are much more diverse.
For instance, fractures can be directly caused by the material decomposition \zp{~\cite{yang2003mechatronic,liu2008thermo,westover2009photoluminescence,zhao2011electrical}} and thermal stress \zp{\cite{song2020progressive,zhao2011electrical}} induced by the Joule heating effects in materials under electrification.
In addition, electromigration, and electrical fatigue can also lead to fractures in metals \zp{~\cite{black1969electromigration,ho1989electromigration,zhao2011electrical}}, and fractures in piezoelectric materials can be caused by fatigue stress under electrical fields\zp{~\cite{yang2003mechatronic}}.

In this work, we investigate electrification-induced fracture in aligned and non-woven CNT films, which exhibit the same 1D nanostructures but radically different microstructures.
The 1D, covalently-bonded graphitic tubules carry the load as well as the heat and charge carriers.
However, the effective thermal diffusion at the film scale is highly anisotropic in the aligned films and isotropic in the non-woven ones, since the van der Waals contacts or gaps between the CNTs are much weaker in energy transfer.
The fracture patterns and their spatiotemporal evolution are characterized to reveal the underlying mechanisms.
We report that the failure is caused by Joule heating that can concentrate at the crack tips, which is supported by evidence from experimental characterization and numerical simulations.
A theoretical model for Joule heating rate at the crack tip and failure criterion is developed in analog to the stress analysis in linear elastic fracture mechanics (LEFM).

\emph{Sample Preparation and Experimental Setup} - 
CNTs are nanofibers with outstanding mechanical performance and thermal conductivity, which can be metallic or semiconducting, depending on the chirality of their graphitic walls\zp{~\cite{dresselhaus1998physical}}.
Preserving the performance of nanofibers in their macroscopic assemblies remains challenging due to the formation of microstructural imperfections.
Efforts have been made to improve the alignment \zp{~\cite{xu2016high,guo2022soft}}, the metallic content \zp{~\cite{ericson2004macroscopic,puchades2015mechanism,guo2022soft}}, and the fill ratio of CNTs in composites \zp{~\cite{han2017preparation,guo2022soft}} in the past decades.
However, the state-of-art techniques are still far from being viable\zp{~\cite{taylor2021improved,wang2021understanding}}.

We fabricate aligned and non-woven films of multiwalled CNTs (MWNTs) with typical diameters of $6-15$ nm\zp{~\cite{jiang2011superaligned}} and lengths of $100-1,000$ \textmu m\zp{~\cite{CHEN201619}}.
Microstructural characterization shows that the MWNTs aggregate into bundles in the aligned and non-woven films, respectively.
The diameters of the bundles ($\approx 30$ nm) also define the thickness of `single-layer' films.
Aligned CNT films are pulled out as continuous ribbons using a tweezer from vertically-grown CNT arrays, the width of which is determined by the size of the initial array\zp{~\cite{jiang2011superaligned}}.
To fabricate non-woven films, CNTs are dispersed in a solvent.
The suspension is then filtered through a porous substrate (a membrane or a filter paper) under controlled vacuum or pressure conditions, where the CNTs form a thin, entangled mat on the substrate, or namely, the non-woven CNT films\zp{~\cite{liu1998fullerene,kim2006fabrication}}.

The aligned CNT films demonstrate highly oriented microstructures in comparison with the non-woven films.
In order to quantify the degrees of alignment of CNTs in the films, a Raman laser beam is directed either parallel or perpendicular to the CNTs.
The recorded intensities of the G band are denoted as $I_{\rm G//}$ or $I_{\rm G\bot}$, respectively, the ratio of which $R = I_{\rm G//}/I_{\rm G\bot}$ provides a direct measure of the orientational order.
We conclude with $R = 2.78$ for anisotropic, aligned films and $R = 1$ for the isotropic, random films.
Improved alignment of CNTs in the films is expected to elevate the stiffness, strength, as well as thermal and electrical conductivities along the strong (fiber) direction\zp{~\cite{gao2018strength,wang2021understanding}}, rendering them with high performance in lightning-strike protection.

\begin{figure}[htb]  
\centering
\includegraphics[width=12cm]{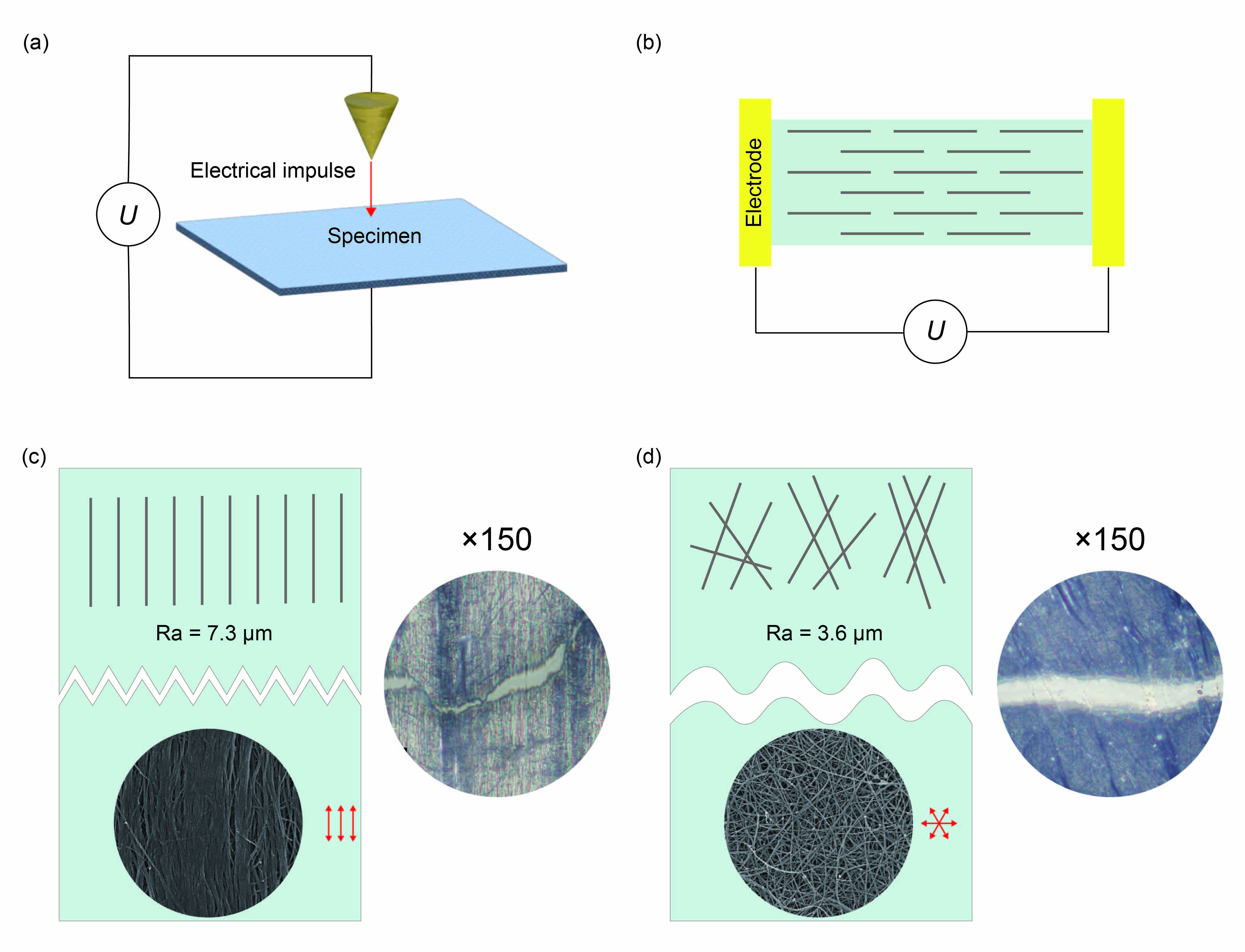}
\caption{
(a) Typical setup of the lighting experiment\zp{~\cite{bai2023superaligned}}.
(b) Our electrical loading device. (c, d) Fiber alignment and optical microscopy (OM) images of fracture patterns in aligned (c) and non-woven (d) carbon nanotube (CNT) films.  
}
\label{fig:1}
\end{figure}

The structural responses and failure of CNT films under lightning strikes are experimentally simulated in an electrified platform, where a quartz substrate structurally supports the films\zfig{~(Fig.~\ref{fig:1}(a, b))}.
Quartz is chosen for its smooth surface, good thermal insulating properties, and thus the least perturbation to the fracture process.
Electrical contacts are made by depositing silver to two opposite edges of films.
Electric fields are then applied by using a direct current (DC) power source (Keithley 2480) with a voltage $V = El$, where $l$ is the length of the film and $E$ is the electrical field.
Experimental tests show that the films are destructed in a time scale of $0.01-0.1$ s, which is longer than the duration of a typical lightning strike\zp{~\cite{larsson2002interaction,kumar2020factors}}.
Post-destruction analysis using optical microscope (OM) images ($1067$ pixels/cm) is used to calculate the edge roughness $R_{\rm a}$.
 The crack morphologies of aligned CNT films are identified to be sharp and jagged with $R_{\rm a}=7.3$ \textmu m \zfig{(Fig.~\ref{fig:1}(c))}.
In contrast, non-woven CNT films are destructed with a diffuse pattern of fracture with much smoother edges ($R_{\rm a}=3.6$) \textmu m \zfig{(Fig. \ref{fig:1}(d))}.

\begin{figure}[htbp]
\centering
\includegraphics[width=12cm]{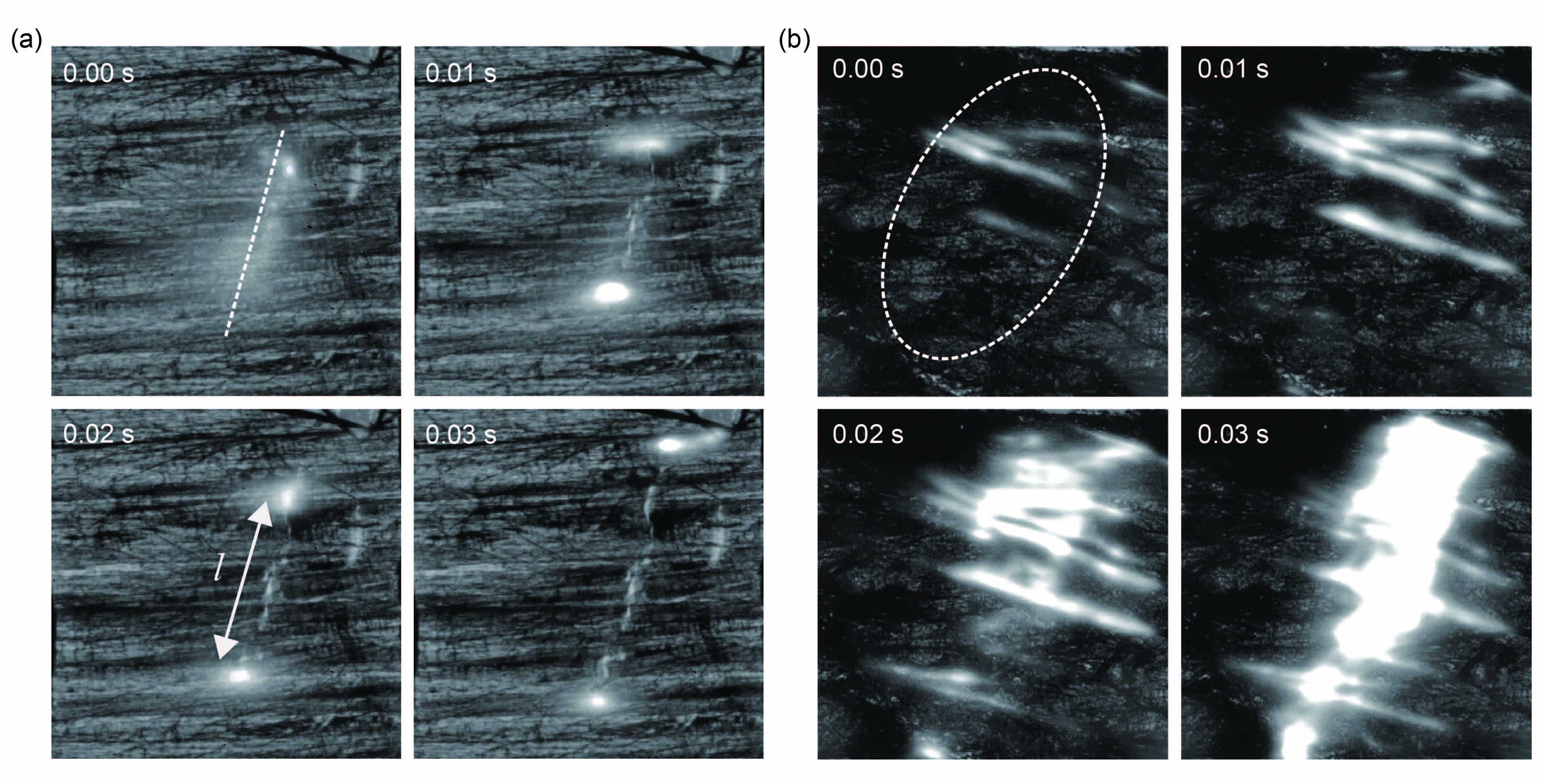}
\caption{
Dynamics of crack propagation in aligned (a) and non-woven (b) CNT films, which show sharp and diffusive patterns under OM. 
}
\label{fig:2}
\end{figure}

\begin{figure}[htb]
\centering
\includegraphics[width=12cm]{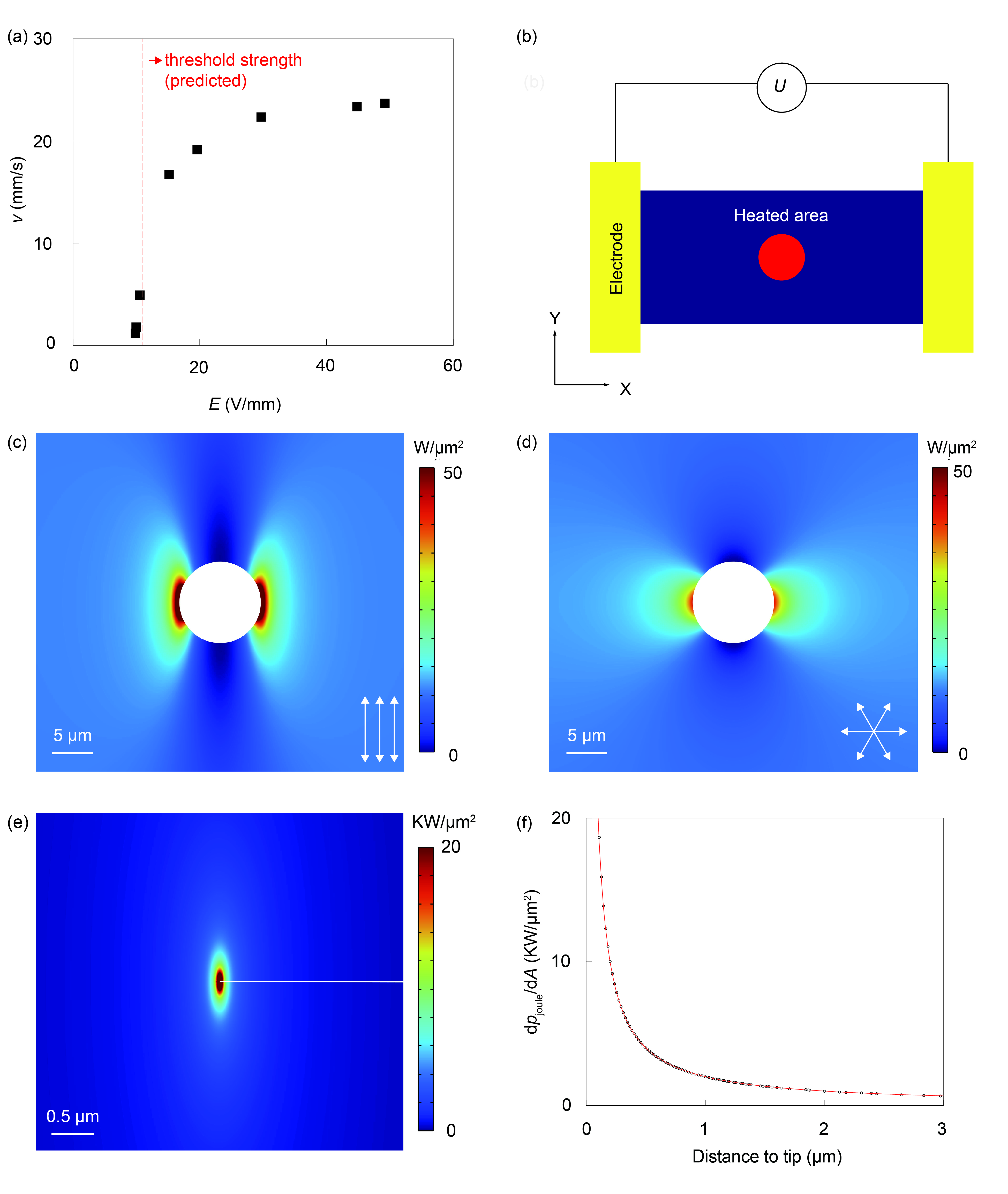}
\caption{
Crack propagation speeds measured in experiments and simulation results of Joule heating rates.
(a) Crack propagation speed $v$ measured under different electrified $E$, the dashed line shows the predicted threshold strength, $E_{\rm c}$ (Eq.~\ref{eq:3}).
(b) The simulation setup.
The electrical field is $E = 40$ V/mm.
The conductivities obtained in our experimental measurements are $40,000$ S/m (fiber direction) $4,000$ S/m (transverse direction) for the aligned CNT films, and $7,000$ S/m for the non-woven films, respectively.
(c, d) Simulation results of the local Joule heating rates (${\rm d}p_{\rm Joule}/{\rm d}A$) in aligned (c) and non-woven (d) CNT films.
(e, f) Local Joule heating rates at a crack tip plotted in the 2D field (e) and as a function of the distance to the crack tip.
}
\label{fig:3}
\end{figure}

Progressive damage of the CNT films is recorded by using a high-speed camera (Photron FastCam mini UX100).
An OM lens ($12\times$) is connected to the camera with a speed of $500$ flashes per second. 
From the videos taken during fracture (\zfig{Fig.~\ref{fig:2}, Supplemental Videos S1 and S2}), we find that in both aligned and non-woven samples, a flash emerges in the center of the specimen at $\sim 0.01$ seconds after power activation.
The occurrence of flash signals energy concentration in the region or localized temperature rise and results in subsequent material damage by fracture.
In the aligned CNT films, the luminous spots are localized at the crack tip and propagate along a straight line in the range of $0-50$ V/mm, leaving a crack wherever it passes (\zfig{Fig.~\ref{fig:2} (a)}).
The speed increases with the applied electric field $E$ after a threshold value of $10$ V/mm and gradually converges to $\approx 24$ mm/s beyond $30$ V/mm\zfig{~(Fig. \ref{fig:3} (a))}.
In contrast, the flash emerges from the central locus and expands uniformly in all orientations in the non-woven CNT films at a speed of $\approx 5$ mm/s, reflecting their isotropic nature.
The fibers along the direction of the electric field show bright spots in the form of line segments, corresponding to the concentration of Joule heat (\zfig{Fig.~\ref{fig:2} (b)}).

\emph{Physics and Theoretical Analysis} -
The coefficients of thermal expansion of CNTs remain as low as $10^{-6}$ K$^{-1}$ even as the temperature rises to $1,600$ K\zp{~\cite{jiang2004thermal}}.
As a result, the maximum thermal stress in the film is far lower than the strength of CNTs, indicating that thermal stress will not lead to fracture.
Buckling of slender CNTs may occur but it will not lead to fracture.
Moreover, CNTs are covalently bonded materials, and electromigration does not play a key role in the destruction process compared to metals\zp{~\cite{bai2023dynamic}}, especially within such a short duration (less than $0.1$ s).
We conjecture that the local damage and destruction are induced by Joule heating.
Balancing the areal rate of Joule heating (${\rm d}p_{\rm Joule}/{\rm d}A$) and that of thermal dissipation ${\rm d}p_{\rm diss}/{\rm d}A$ per unit area $A$ thus predicts the threshold field strength, $E_{\rm c}$.
Beyond $E_{\rm c}$ heat will be accumulated and fracture of the film is activated as observed in our experiments.
Typically the failure will be nucleated at the center of the film\zp{~\cite{liu2016continuous}} due to the boundary effect and at the defective sites with poor conductivities.

Following Joule's law, the rate of heating is

\begin{equation}
{\rm d}p_{\rm Joule}/{\rm d}A=E^{2}t/\rho,
\label{eq:1}
\end{equation}

\noindent where $t$ and $\rho$ are the thickness and electrical resistivity of the CNT films, respectively.
On the other hand, heat dissipation of CNT films includes contributions from natural convective heat dissipation, $p_{\rm C}$, and radiative heat dissipation, $p_{\rm R}$, that is

\begin{equation}
{\rm d}p_{\rm diss}/{\rm d}A=p_{\rm C}+p_{\rm R}=h_{\rm c}(T-T_{\rm a})+\epsilon\sigma(T^4-T_{\rm a}^4),
\label{eq:2}
\end{equation}

\noindent where $h_{\rm c}$ is the natural convective heat transfer coefficient.
$T_{\rm h}$ and $T_{\rm a}$ are the temperatures of sample surfaces and ambient air, respectively.
The radiative heat dissipation is determined by the Stefan-Boltzmann law, where $\epsilon$ and $\sigma$ are the surface emissivity ($0<\epsilon<1$), and the Stefan-Boltzmann constant ($5.67\times10^{-8}$ W$\cdot$m$^{-2}$K$^{-4}$).
A threshold field strength can thus be estimated as

\begin{equation}
E_{\rm c}=\sqrt{(\rho h_{\rm c}(T_{\rm b}-T_{\rm a})+\epsilon\sigma(T_{\rm b}^4-T_{\rm a}^4))/t},
\label{eq:3}
\end{equation}

\noindent beyond which the Joule heating rate ${\rm d}p_{\rm Joule}/dA$ becomes higher than the maximum value of ${\rm d}p_{\rm diss}/{\rm d}A$.
Considering the breakdown (oxidization) temperature of CNT films in the air as $T_{\rm b} = 820$ K\zp{~\cite{bai2023dynamic}}, the estimated value of $E_{\rm c}$ shows agreement with the experimental data\zfig{~(Fig.~\ref{fig:3}(a))}.

Above $E_{\rm c}$, the crack initiates from the destructed region (the central hole in \zfig{Fig. \ref{fig:3}(b)}) and starts to propagate at a speed increasing with the Joule heat injected into the crack front.
The heat concentration near the crack tip is simulated by finite element analysis (FEA).
A 2D model with anisotropic thermal and electrical conductivities is constructed for the CNT films, which can be reduced to the isotropic one.
The initial damage is modeled as a circular hole in the sample.
The simulation results show the concentration of Joule heating near the damaged area \zfig{(Fig.~\ref{fig:3}(c,d))}.
Compared with the isotropic, non-woven CNT films, there is a more intense concentration in the vicinity of the initial damage in aligned CNT films.
The crack in the aligned CNT films thus propagates along a straight line, while in the non-woven film, the crack expands in a diffuse way.

By considering the anisotropic electrical conductivity in the aligned CNT films, the electrical potential $\phi$ can be solved from the governing Poisson equation

\begin{equation}
\sigma_{x}\frac{\partial^2}{\partial x^2}\phi+\sigma_{y}\frac{\partial^2}{\partial y^2}\phi=0.
\label{eq:4}
\end{equation}

By rescaling the coordinates $x'=x/\sqrt{\sigma_{x}/\overline{\sigma}}, y'=y/\sqrt{\sigma_{y}/\overline{\sigma}}$, where $\overline{\sigma}=(\sigma_x+\sigma_y)/2$ is the average conductivity, this equation can be transformed into $(\frac{\partial^2}{\partial x'^2}+\frac{\partial^2}{\partial y'^2})\phi=0$.
The boundary conditions of electrically insulating edges are enforced.
The near-field field strength in polar coordinates is obtained as  \zp{~\cite{hua2005crack}}
\begin{equation}
E_x=-E\sqrt{\frac{a}{2r}}{\rm sin}\frac{\theta}{2},\\
E_y= E\sqrt{\frac{a}{2r}}{\rm cos}\frac{\theta}{2},
\label{eq:5}
\end{equation}
\noindent where $r=\sqrt{x'^2+y'^2}$ measures the distance to the crack tip.
$\theta$ and $a$ are the corresponding angle and crack length in the transformed coordinates. 
This result features a similar $-1/2$ singularity as the stress field at the crack tip in LEFM.
As shown in the coordinate transformation, the anisotropic conductivity enhances the field concentration at the crack tip.
The rate of Joule heating in the vicinity of the crack tip thus exhibits a direct proportionality to the inverse of the distance from the crack tip, signaling heat concentration \zfig{(Fig.~\ref{fig:3}(e, f))}.
In practice, the electrified at the crack tip cannot go to infinity due to the shielding effect originating from charge redistribution in the CNTs.
An increase in $E$ will result in a higher heating rate, ${\rm d}p_{\rm Joule}/{\rm d}A$ and a more rapid local destruction process \zfig{(Fig.~\ref{fig:3}(a))}, which saturates as the crack approaches the sample boundary.

\begin{figure}[htb]
\centering
\includegraphics[width=12cm]{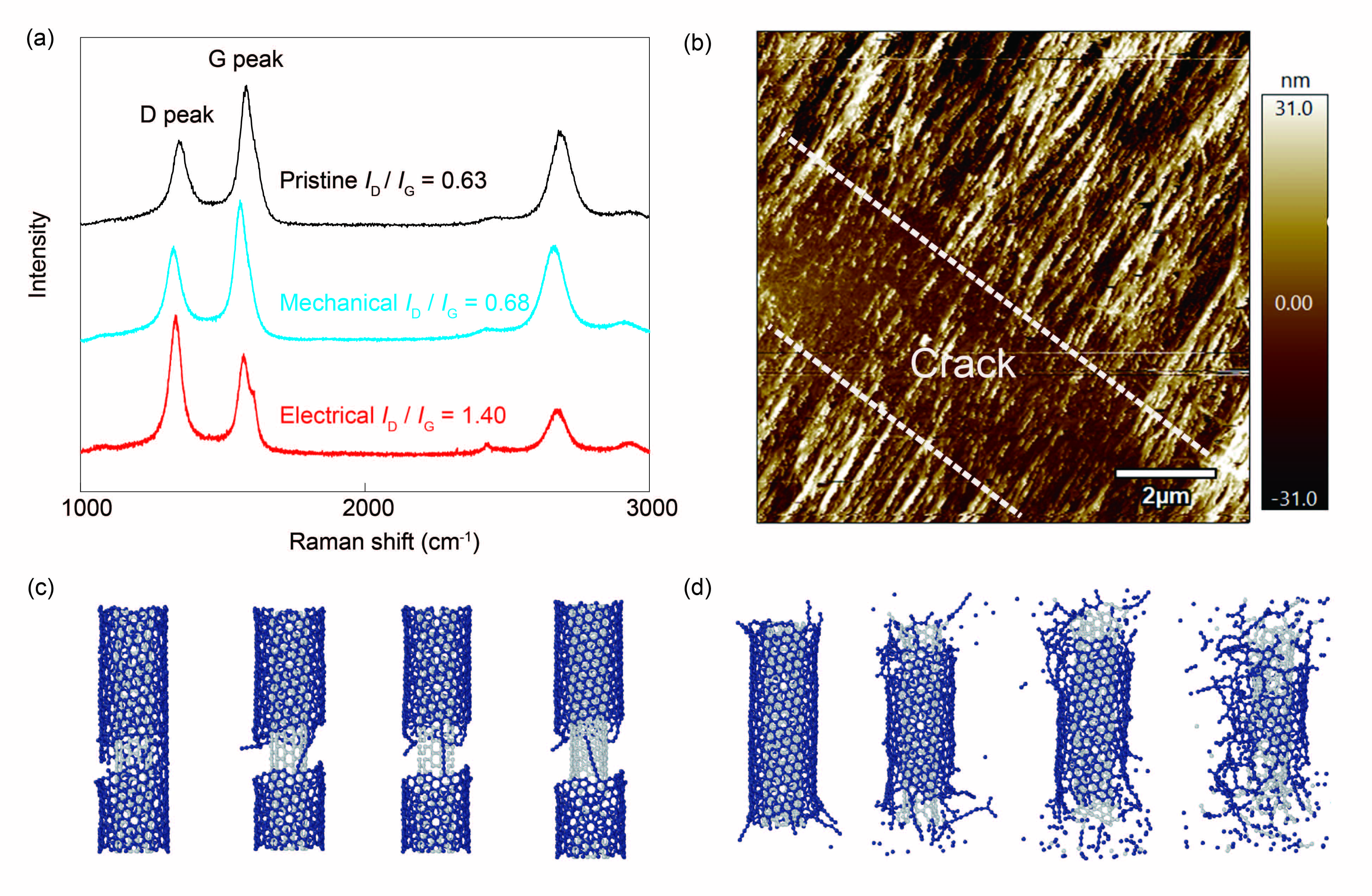}
\caption{
 (a) Raman intensities of pristine, mechanically, and electrically damaged aligned CNT films.
 (d) Atomic force microscopy (AFM) image of the crack zone.
 (c, d) Molecular dynamics (MD) simulation snapshots of the ($6, 6$)@($11,11$) double-walled CNTs (DWNTs) damaged by tensile stress (c) and Joule heating (d).
}
\label{fig:4}
\end{figure}

The fractured surfaces of aligned CNT films under electrification are further examined by Raman spectroscopy to elucidate the atomic-level processes occurring at the crack tip, and compared to the surfaces of cleaved mechanically \zfig{(Fig.~\ref{fig:4}(a))}.
The results suggest a higher intensity of the $I_{\rm D}/I_{\rm G}$ ratio and thus a greater degree of atomic-scale disorder in the electrified samples.
Atomic force microscopy (AFM) images at the crack tip resolve the crack features created by Joule heating, which show jagged patterns \zfig{(Fig.~\ref{fig:4}(b))} and suggest a melting failure at the single-fiber level under electrification.
We also carry out molecular dynamics (MD) simulations for direct evidence of these processes.
Tensile- and electrification-induced fractures of double-walled CNTs (DWNTs) are modeled using the large-scale atomic/molecular massively parallel simulator (LAMMPS) \zp{~\cite{thompson2022lammps}} \zfig{(Fig.~\ref{fig:4}(c, d))}.
The adaptive intermolecular reactive empirical bond order (AIREBO) potential \zp{~\cite{stuart2000reactive}} is used to model the interatomic interaction between carbon atoms in the DWNTs.
Periodic boundary conditions are used with a supercell of $40\times40\times100$ nm.
Joule heating is performed by raising the temperature from $350$ K to $3,500$ K at a rate of $3.2$ K/ns, and the strain rate of tensile loading is $2$ m/s, which is relatively low to reduce the rate dependence but still much higher than the experimental values due to the spatiotemporal limitation of MD simulations.
The findings indicate that during tensile fracture, the DWNTs experience bond breakage, while the sp$^{2}$ lattice remains unaltered.
In contrast, the lattice melts under Joule heating. 

In this study, we conduct a comparative analysis of the fracture patterns shown by aligned and non-woven CNT films.
In terms of fracture edges, aligned CNT films exhibit a distinct and irregular morphology, characterized by sharp and jagged features.
In contrast, the edges of the non-woven CNT films appear to be more refined and less rough.
Aligned CNT films demonstrate a linear, tip-guided fracture propagation mode, whereas non-woven CNT films display a diffusive mode.
Numerical simulations and theoretical analysis show that the observed discrepancy can be attributed to the pronounced anisotropic conductivity of the aligned CNT films.
This characteristic gives rise to a significant heat accumulation around the edges of the initially damaged hole in the aligned CNT films.
We then theoretically demonstrated the existence of a singular temperature field near the crack tips, which guides the dynamics of cracking and defines the fracture patterns.
The arguments exhibit a favorable agreement with the outcomes obtained from the experimental and simulation results.
The fracture patterns of the aligned CNT films are further examined by Raman spectroscopy, AFM characterization, and MD simulations, validating that the fracture in electrified films is generated by Joule heating and is very different from force-induced fracture.

This study was supported by the National Natural Science Foundation of China
through grants 52090032 and 11825203, and the National Key Basic Research Program of China grant No. 2022YFA1205400.
The computation was performed on the Explorer 100 cluster system of the Tsinghua National Laboratory for Information Science and Technology.

\bibliography{main_text}

\clearpage
\newpage

\clearpage
\newpage

\clearpage
\newpage

\clearpage
\newpage

\clearpage
\newpage

\begin{figure}[htb]
\centering
\includegraphics[width=8cm]{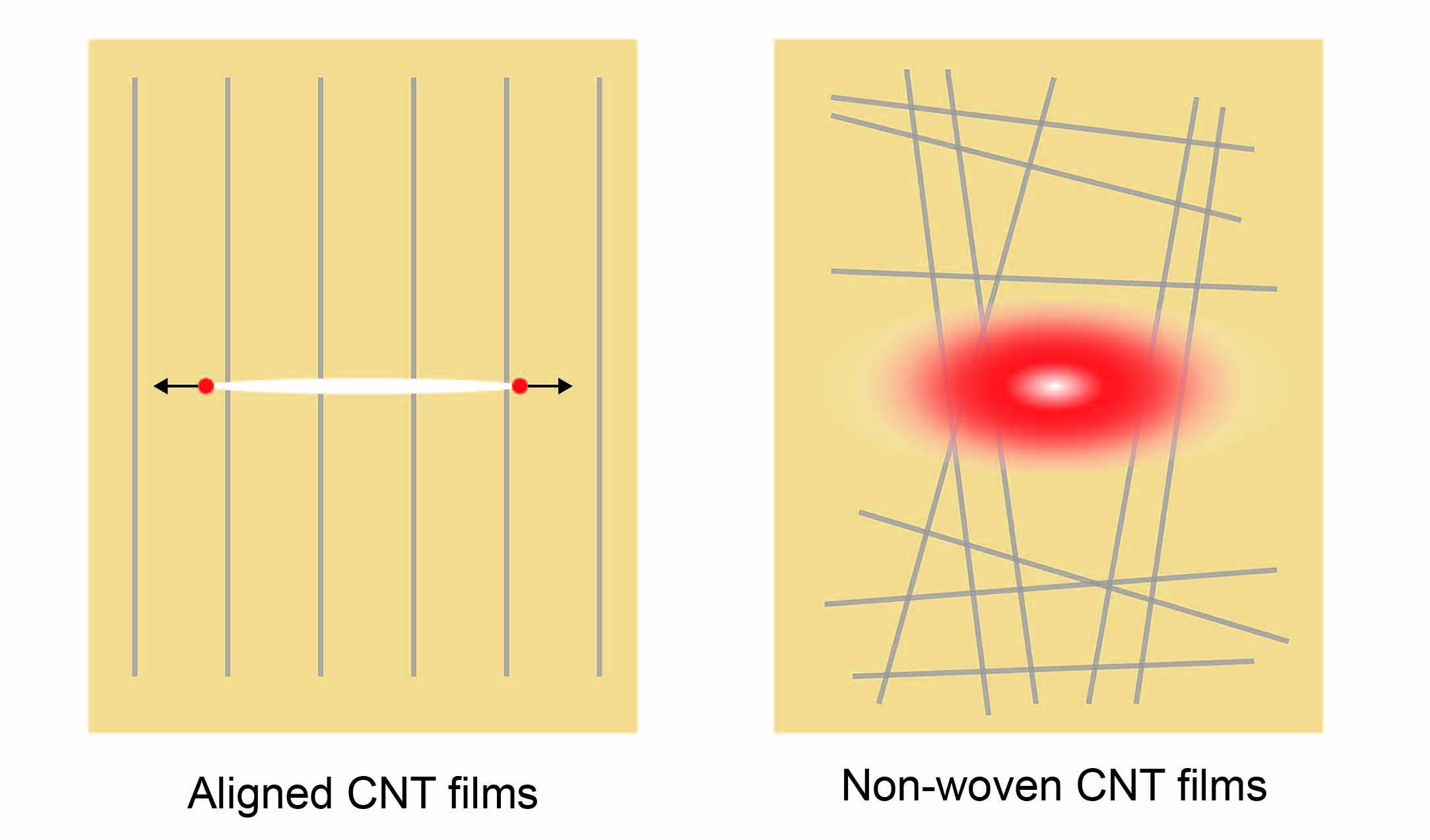}
\caption*{Table of Content Graphics}
\end{figure}

\end{document}